\documentclass{article}
\usepackage[utf8]{inputenc}
\DeclareUnicodeCharacter{2212}{-}
\usepackage[letterpaper, margin=1in]{geometry}
\usepackage{graphicx}
\usepackage[section]{placeins}
\usepackage{subcaption}
\usepackage{float}
\usepackage{multicol}
\usepackage{multirow}
\usepackage{amssymb}

\usepackage[sorting=none]{biblatex}
\usepackage[table]{xcolor}

\def\ee{e$^+$e$^-$}
\def\tt{$t{\bar t}$}

\newcommand\snowmass{\begin{center}\rule[-0.2in]{\hsize}{0.01in}\\\rule{\hsize}{0.01in}\\
\vskip 0.1in Cross-Frontier Report Submitted to the US Community Study\\ 
on the Future of Particle Physics (Snowmass 2021)\\ 
\rule{\hsize}{0.01in}\\\rule[+0.2in]{\hsize}{0.01in} \end{center}}

\title{\snowmass\textbf{Report of the Snowmass 2021 e$^+$e$^-$-Collider Forum}}
\author{Maria Chamizo Llatas, Sridhara Dasu, Ulrich Heintz,\\
Emilio A. Nanni, John Power, Stephen Wagner}
\date{\today 
}

\begin{document}

\maketitle
\newpage
\tableofcontents
\newpage

\section{Executive Summary}\label{exec}

Electron-Positron colliders have the potential to span particle physics from precision electroweak measurements to discoveries at the Energy Frontier. Future \ee Higgs Factories -~linear and circular~- are capable of providing a rich scientific program addressing a broad range of fundamental topics at the Energy Frontier, FCC-ee~\cite{FCC-ee}, CEPC~\cite{CEPC}, CLIC~\cite{CLIC}, ILC~\cite{ILC}, with sub-percent Higgs boson coupling measurements, and potentially discovering the next new physics scales up to O(10) TeV. The near term plan includes high precision electroweak measurements of the Z, W, top and Higgs to study deviations from the Standard Model, including top quark and triple Higgs couplings. This program is possible with current technologies and will advance our understanding well beyond the HL-LHC. In the long-term, the central challenge for the \ee~collider community is to provide an upgrade path to the Energy Frontier while mitigating cost. Both the linear and circular long-term upgrade paths are based on reusing the investment in civil engineering of near-term and mid-term machines to reach parton energies of $O(10)$ TeV to significantly broadening the scientific program.

Circular colliders will be implemented in stages running at the Z, WW threshold, ZH, tt pair production. The ultimate upgrade path is a follow-on hadron collider, which is outside the scope of this report. Near term linear colliders provide most of the statistics at the ZH, and will also run at Z, WW and tt threshold, providing polarized electrons and positrons to enhance the signal.  Linear colliders provide an upgrade path for energies above 0.5 TeV. Long-term linear collider proposals aim to lower cost by increasing acceleration gradient and lowering power consumption. The new $\rm{C}^3$ \cite{CCC-Snowmass} concept has made progress on both fronts. Very-long term options will require significantly more accelerator R\&D but will dramatically increase gradient ($>$1~GeV/m) and efficiency with Wakefield Accelerators (WFAs) with strawman designs starting at $1$-TeV and the potential to reach the $O(10)$-TeV scale, or Energy Recovery Linacs (ERL) to reduce power consumption while providing very high luminosities and center of mass energy, such as CERC~\cite{CERC}, ReLiC~\cite{ReLiC} and ERLC.

A circular Higgs Factory will provide the best precision for most Higgs couplings, but direct probing of Higgs self-coupling and ttH couplings is deferred to a future higher energy proton collider. Whereas a linear Higgs Factory will provide access to the Higgs self-coupling and ttH coupling. 

The primary consideration for the delivery of physics results is the start time of the physics program. Given the maturity of the technology, the ILC holds the advantage for an early start of the program. The FCC-ee and CEPC are able to complete the required runs at various luminosities faster but their larger civil engineering work requires significantly more time and cost. An early start of the civil engineering construction of a circular machine is therefore key to timely realization of physics. The ILC and $\rm{C}^3$ have cost, higher energy-reach, and polarization advantages but with lower luminosity, needing significantly longer running time to achieve the same level of precision for measurements compared to circular machines. Among the newer proposals only $\rm{C}^3$ proposes a timescale which is suitable for early physics, although it does require an early demonstrator. From a potential siting point of view all but the $\rm{C}^3$ machine require green-field sites. Development of WFA-based $O(10)$-TeV scale machine, with sufficient luminosity capability for O(10)~ab$^{-1}$, and energy-recovery technologies for improved power-to-luminosity costs, requires continued R\&D investment.

In order to address the Detector R\&D and preparation of a Technical Design an R\&D program that goes beyond generic R\&D is needed to address the specific challenges posed by the detector required for \ee colliders. Such a program needs to start now for the technology to build a full scale \ee collider detector to be ready when the HL-LHC program is completed. 

The vast majority of $\sim 30,000$ currently  operating accelerators globally are electron accelerators. Electron accelerator R\&D ranges from industrial applications to the cutting edge development of ultimate storage rings and linear accelerator based XFELs. This fortunate situation allows \ee colliders to leverage these global efforts to provide a viable path to a collider reducing the R\&D costs to the HEP budget.

Given the strong motivation and existence of proven technology to build an \ee ~Higgs Factory in the next decade, the US should participate in the construction of any facility that has firm commitment to go forward. Awaiting such commitment, the US should also pursue research and development of multiple options in this decade. This ensures that the global community will be able to begin constructing at least one such machine in the following decade. Potential siting of a facility in the USA should also be pursued. US investments in further advancing technology will ensure the technical readiness of proposed facilities, improves the eventual physics reach of a collider and maintains the community engagement needed for the US to contribute to the construction of a collider.

\pagebreak

\section{Introduction}\label{intro}

Discovery of the 125-GeV Higgs boson~\cite{Higgs-ATLAS, Higgs-CMS} and the non-appearance of new particles at the TeV scale, as anticipated in well-motivated Beyond the Standard Model scenarios, puts us at crossroads. While the discovery of the Higgs boson was a formidable success of the Standard Model, it leaves a number of questions unanswered, and in fact marks the opening of a new and exciting era of exploration. While there is plenty of parameter space left to explore at the HL-LHC with advanced techniques, we must plan for future exploratory machines now. Some of the points to be addressed at future facilities include the measurement of the Higgs couplings to all charged fermions, connections between the Higgs field and the origin of the neutrino masses, existence of additional Higgs bosons, supersymmetry, or the study of the electroweak symmetry breaking through the measurement of the Higgs self-couplings. Searches for new signatures that could explain the origin of dark matter, or the origin of the neutrino masses, will require complementarity between \ee ~colliders and very high energy colliders or complementarity with other direct searches facilities.

While no new results are presented here, the extracts of the summaries of various documents submitted to the Snowmass community study were used to provide this report to assist in the preparation of the energy frontier report. In this report, we considered the physics case and the proposals of the major \ee ~collider efforts that are currently under consideration by the Snowmass community, using the detailed comparisons provided by the Integrated Task Force (ITF). This survey is divided into 3 energy sectors, $<$ 1 TeV, $1-3$ TeV, and $O(10)$ TeV and we summarize the \ee ~collider efforts by their physics capabilities, maturity, and readiness to deliver the scientific program in the near future as well as the potential path for upgrades. 

A majority of theoretical papers conclude that percent-level precision of the Higgs coupling measurements is necessary to derive useful constraints on new physics contributions. With percent-level deviations from the Standard Model expectations in the Higgs sector or the electro-weak sector in general, a new physics scale of approximately 10-TeV could be established.

Building an \ee-Higgs factory as the next frontier machine to make precision 1\%-level measurements of Higgs boson couplings is broadly recognized as the next important goal of a new particle physics machine. Such an exploration of the Higgs sector is likely to point to the 10-TeV regime as the next energy frontier of interest. Therefore, the desire of the community is to find an optimal strategy to achieve a Higgs Factory, soon followed by an energy frontier machine at the 10-TeV partonic scale.

\begin{table*}[ht!] 
    \centering
		\caption{\label{table:ee-colliders} \ee accelerators submitted to the Snowmass process.}
	\begin{tabular}{c | c  }
		{Collider} & {Type}   \\
		\hline
		FCC-ee (0.24 TeV) & Circular \\
        CEPC (0.24 TeV)    & Circular  \\
    	CERC (0.24 TeV) & Circular\\
    	\hline
        ILC (0.25 TeV) & Linear  \\
		CLIC (0.38 TeV) & Linear  \\
        CLIC (3 TeV) & Linear  \\  
		C$^3$ (0.25 TeV) & Linear   \\
		ReLiC (0.24 TeV) & Linear  \\
		ERLC (0.24 TeV) & Linear \\
		ILC (3 TeV) & Linear \\
		C$^3$ (3 TeV) & Linear \\  
		ReLiC (3 TeV) & Linear \\
		WFA (3 TeV) & Linear \\
		WFA-flat (15 TeV) & Linear \\
		WFA-round (15 TeV) & Linear  \\
	\end{tabular}\\
\end{table*}
\section{Physics }\label{phys}

This document summarizes goals, plans and capabilities of a plethora of \ee-colliders proposed (See Table \ref{table:ee-colliders}) for the future studies of high energy physics. The proposed circular \ee-colliders, FCC-ee and CEPC are 100-km class machines which allow high luminosity operation as Higgs Factories extensible through \tt-threshold, and include a second phase as hadron colliders at around 100 TeV, whereas the linear \ee-colliders, ILC, CLIC and C$^3$ with varying degrees of maturity, begin as Higgs factories and are capable of TeV-scale extension. Other options for linear and circular \ee ~colliders based on energy recovery Linac technology have been proposed, like CERC, ReLiC, ERLC. The \ee-colliders based on these advanced technologies target multi-TeV scale operation. 

This section discusses briefly the capabilities of various proposed \ee ~colliders organized by physics topic. Where no collider-specific contribution is available the closest similar collider is cited to glean a ``reasonable'' estimate. The Snowmass studies have been performed using the running scenarios tabulated in Table \ref{fig:Run-scenarios-table}. 

\begin{table}[H]
   \centering
    \includegraphics[width=3in]{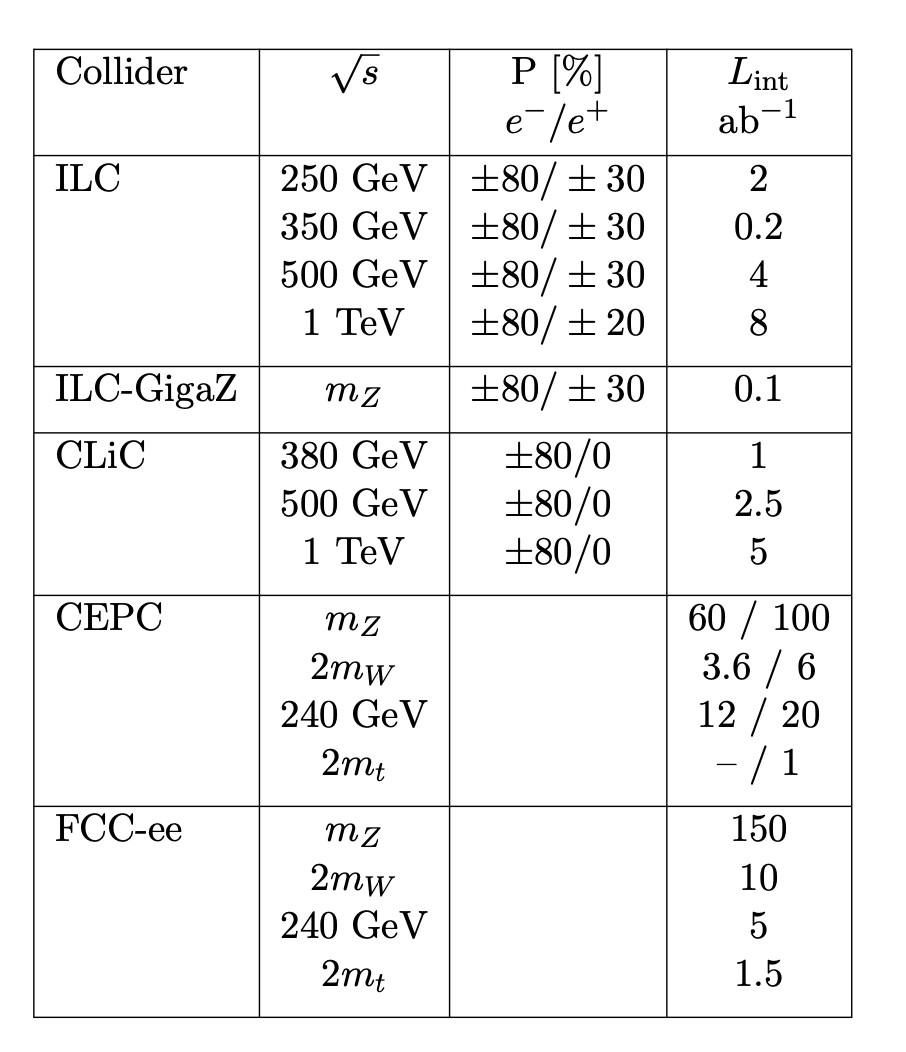}
    \caption{Running scenarios used for the Snowmass 21 studies.}
    \label{fig:Run-scenarios-table}
\end{table}

\subsection{Higgs and Electroweak Machines: ($<$ 1 TeV)}

While there are several center-of-mass energy and luminosity choices, the most mature case for an \ee-machine is that of the Higgs Factory with a center of mass energy which is able to produce the ZH process with luminosities between $10^{34}$ and $10^{35}$ cm$^{-2}$s$^{-1}$. The Higgs self-coupling measurement requires higher energies: around 500-GeV energy for the ZHH process and potentially also 1-TeV to access the W-fusion production mode. 

At a 250-GeV lepton collider, the ZH process provides a clean sample of Higgs bosons. The collider center-of-mass energy constraint allows measurement of all decays of the Higgs boson, essentially independent of the decay final state, using the Z-recoil measurement. While theoretical uncertainties require further reduction, the experimental systematic uncertainties are small compared to statistical uncertainties. Therefore, the collider with the highest integrated luminosity is favored, when neglecting its cost and construction timeline. Polarization of the electron beams can result in improved uncertainties somewhat compensating for lower luminosity.

Operation at various threshold energies where interesting physics processes turn-on, e.g., $WW$-production, $t\bar{t}$-production, $ZHH$-production, is of significant interest. These processes allow precise measurement of the W-mass, which is of significant interest lately, Higgs-top coupling or the Higgs self-coupling.

Proponents of FCC-ee, CEPC, ILC, CLIC and C$^3$ have outlined operations at various centers of mass energies. The threshold scan at the $WW$-threshold to determine the W-boson mass to MeV-precision is of interest noting the new results from the CDF collaboration. Similarly, the $t\bar{t}$-threshold scan and measurements can be done for the first time at an \ee-facility to obtain improved understanding of the top-quark properties. 

Of particular interest is $ZHH$-production, which provides a window into the Higgs-boson self-coupling. However, the low cross section for the process only allows 10\%-level measurement. Direct access to this process could be achieved, provided high enough center of mass energies can be reached with \ee ~colliders with decent luminosities (ILC-500, C$^3$, CERC, ReLiC, ERLC and plasma wakefield colliders), with multi-TeV hadron colliders (FCC-hh, SppS), or with muon colliders. The varying degree of maturity of some of these concepts requires a significant R\&D program to achieve the energy and luminosities quoted in the papers.  


\subsubsection{Higgs \& Electroweak precision physics - up 365 GeV}

The two most important production processes to measure the Higgs couplings for energies below 365 GeV are  Higgsstrahlung (ZH), \ee $ \rightarrow$ ZH, and WW fusion to a Higgs boson (WWH), \ee $ \rightarrow$ H$\nu_e\overline{nu}_e$. Given the cross sections of these processes and the running scenarios in which large data samples will be collected, precision measurements of the Higgs cross sections, decay width and mass will be possible. The statistical precision quoted by FCC-ee for the total ZH cross section is at the per mille level. The expected uncertainties of the Higgs branching fraction to bb, cc, gg, W$^+$W$^-$, ZZ, $\tau^+\tau^{-}$, and invisible, have been estimated to be at the percent level and below, using the ZH and WWH production modes at different center of mass energies. 

At these energies precise measurements of the Higgs branching fractions are possible, and the measurement of the Higgs boson couplings to the Standard Model particles, where the precision for a large set of couplings is expected to reach below the 1$\%$ level. The measurements of H$\mu\mu$ and H$\gamma\gamma$ couplings, and the Higgs self-coupling will reach the ultimate precision either with hadron machines, or with higher \ee~collider energies.  The Higgs self-coupling can be be constrained at low-energy \ee colliders via loop corrections to single Higgs boson processes. 

Figure~\ref{fig:higgs-couplings-comparison} tabulates the precision for the Higgs couplings that can be reached with different machines at different center of mass energies and integrated luminosities. Circular colliders provide higher luminosities and need less time to achieve the ultimate precision (FCC-ee) or can run longer to improve the precision (CEPC). 

\begin{table}[H]
   \centering
    \includegraphics[width=6in]{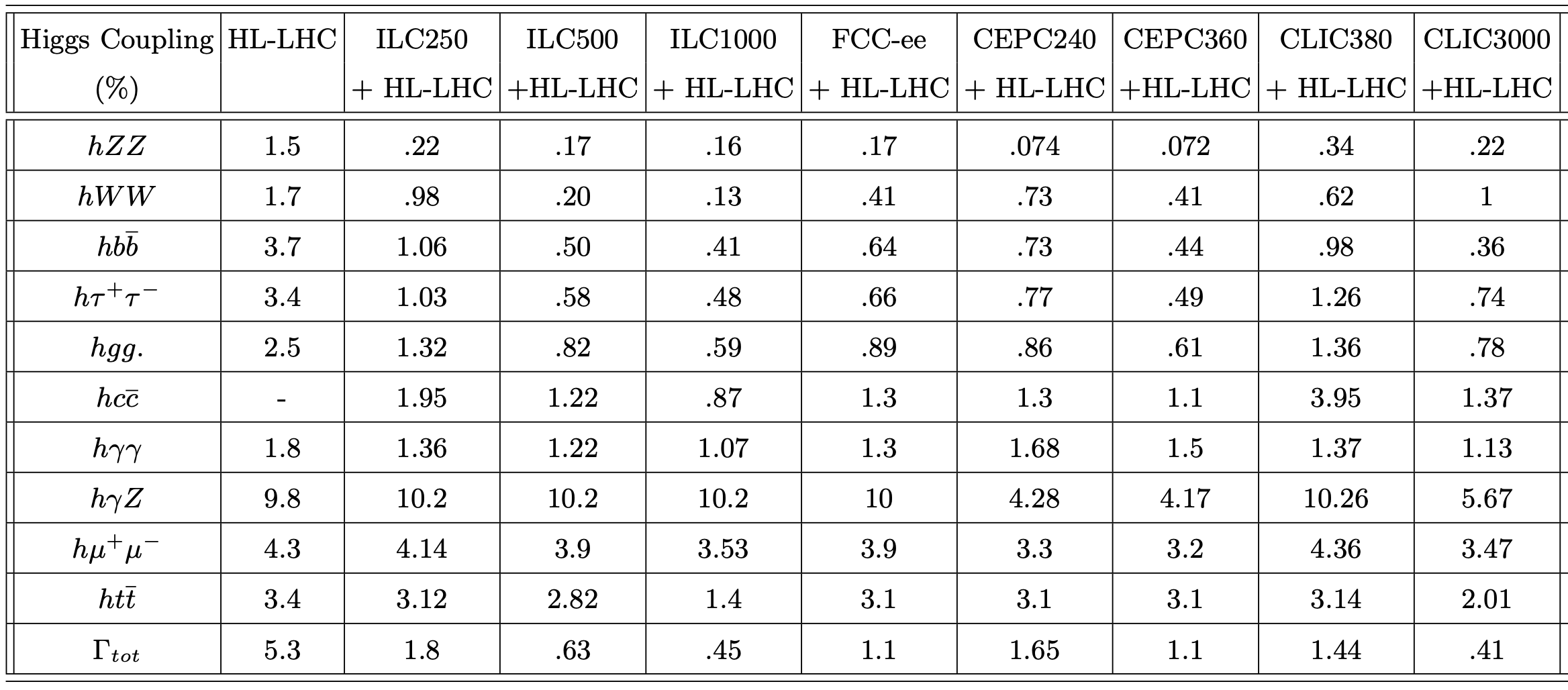}
    \caption{Precision for the Higgs couplings obtained with different accelerators. Extract of \ee-collider entries of Table XX from the Higgs Topical group.}
    \label{fig:higgs-couplings-comparison}
\end{table}

Precision electroweak measurements at \ee~colliders will constitute an important part of the physics program, with a sensitivity to new physics that is very broad and largely complementary to that offered by measurements of the Higgs boson properties and flavor observables. The combination of large data samples at different center-of-mass energies from the Z to above the top quark pair threshold and continuous parts-per-million control of the beam energy at the Z and WW threshold will allow the experimental precision of many electroweak precision observables to be improved by 1–3 orders of magnitude with respect to current measurements. 

For “canonical” electroweak precision measurements (Z-pole, WW threshold), circular \ee colliders (FCC-ee, CEPC) have in general a higher sensitivity than linear colliders (ILC, CLiC) due to the high luminosity at center-of-mass energies below 200 GeV (tabulated in Table \ref{fig:EPO-EW-observables}). Beam polarization at the linear colliders improves their sensitivity and can help control systematic uncertainties.

\begin{table}[H]
   \centering
    \includegraphics[width=6in]{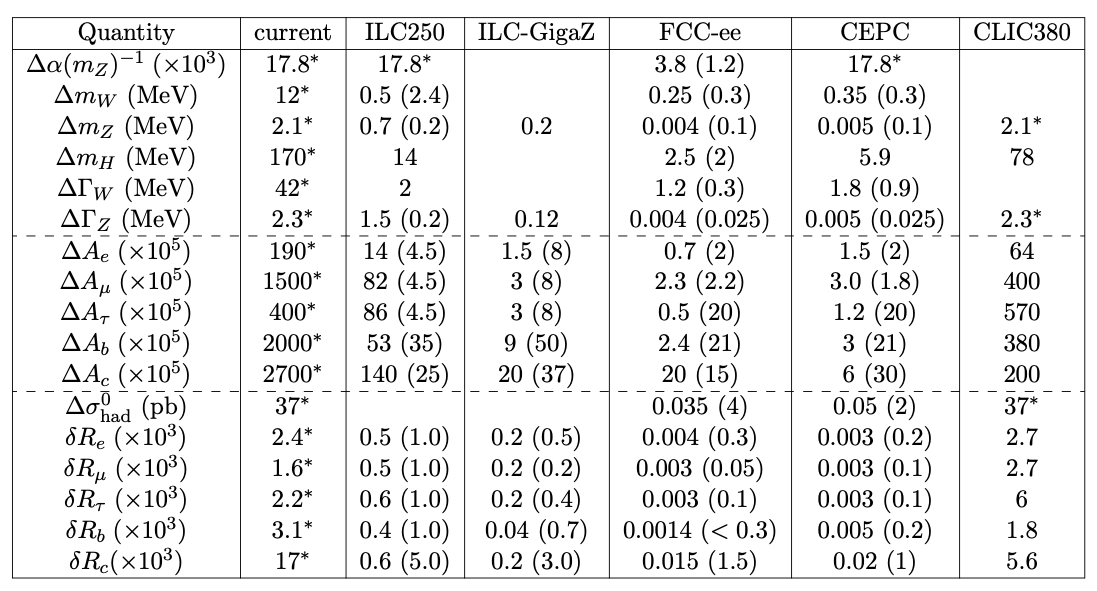}
    \caption{Uncertainties in the Electroweak Precision Observables with different \ee colliders, statistics (systematic). Table 3 from Electroweak Topical Group report.}
    \label{fig:EPO-EW-observables}
\end{table}

ILC, FCC-ee and CEPC propose an energy scan around the tt threshold production to measure the top quark mass using integrated luminosities of 100-200~fb$^{-1}$ with the expectation to reach a statistical uncertainty down to around 10 MeV depending on the number of free parameters in the fit and the number and range of energy points. The systematic uncertainty due to the strong coupling constant adds up to approximately 50 MeV, with the current state-of-the-art calculations and world average for $\alpha_s$. 

 Both ILC and FCC-ee propose a full program to measure the top electroweak Yukawa, $tt\gamma$, and $ttZ$ couplings at around the $t{\bar t}$ threshold and provide sensitivity to anomalous vector coupling of $\gamma(Z)$ of around $1(3). 10^{-3}$ (FCC-ee). 

Flavor-changing neutral current (FCNC) interactions constitute an excellent tool for constraining the SM and probing new physics in the top quark sector. FCNC interactions in the top quark sector involve the exchange of a neutral boson (H, $\gamma$, g, Z) rather than a W boson. The studies for Flavor Changing Neutral Currents of the top quark performed by ILC and FCC-ee and show that at an \ee collider the branching fractions can improve the LEP/HL-LHC sensitivity by one to two orders of magnitude. ILC studies show that polarization will also help to improve precision.  

Understanding QCD physics is key to reducing the systematic experimental and theoretical uncertainties for an enormous variety of observables. Precise knowledge of the strong force is a prerequisite to properly interpret all collider measurements in precision SM studies and BSM searches. The precision for the strong coupling constant, $\alpha_s$, quoted by FCC-ee with the statistics collected at the Z peak reaches 0.1$\%$. The clean environment of \ee colliders will also provide enormous (multi)jet data samples to improve our understanding of parton showers, higher-order logarithmic resummations, as well as hadronization and non-perturbative phenomena.   

The runs at the Z peak will provide unprecedented samples of Z bosons decaying to b-quarks, c-quarks and $\tau$-leptons enriching the knowledge of the flavor physics of quarks and leptons both quantitatively and qualitatively beyond the LHCb and Belle-II programs. The huge sample of b-flavored hadrons in an experimentally clean \ee environment combined with the large boost from the decay of the Z boson will enable a rich b decay program at FCC-ee with unique sensitivities. Measurements of rare heavy quark decays, precise measurements of CKM matrix elements and CP-violating processes can not only strengthen our understanding of the SM but also indirectly probe new physics at scales much above the beam energy, differently compared to those probed by Higgs/EW precision measurements. Some examples like the measurements of leptonic and semileptonic flavor changing neutral current decays provide tight constraints on new physics. The vast amount of data will also allow measurements of the CP-violation parameters and CKM angles, charged lepton violation decays in the Higgs, Z, and $\tau$ decays. An additional feature will be the study of the $\tau$ lepton properties, like $\tau$ lifetime and leptonic branching fractions with very high precision.

The smallest Yukawa coupling in the SM (with zero-mass neutrinos) is that of the electron. Measuring the Higgs coupling to the electron is impossible at hadron colliders because of its tiny branching fraction and the large Drell-Yan background. FCC-ee proposes a dedicated run at $\sqrt s$=m$_H$, which requires the beams to have a very small energy spread, to access the electron Yukawa coupling.

Other measurements proposed at ILC at 250 GeV is to prove the CP nature of the Higgs boson.

\subsubsection{Direct access to Higgs self-coupling and top Yukawa couplings ~600 GeV - 1 TeV} 

At center of mass energies above 500 GeV, the self-interaction of the Higgs boson, in particular the triple-Higgs coupling, can be proved directly by studying the production of Higgs boson pairs. The two relevant di-Higgs production processes are double Higgs-strahlung (for energies around 500-600GeV), \ee $ \rightarrow ZHH$, and di-Higgs production in WW fusion, \ee $ \rightarrow$ HH$\nu_e\nu_e$,   important at and above 1~TeV in \ee colliders. 

Recent ILC studies show that a 21-22\% precision could be reached for the triple Higgs couplings with improved b-tagging and when adding the $\rm{H H} \to \tau^+\tau^{-} \rm{b\overline{b}} $ to the previous studies. A precision on the triple Higgs coupling down to 10\% could be reached when adding 8 ab$^{-1}$ at 1 TeV.  

A suitable choice of e+ and e- polarization enhances the cross-sections for double Higgs production in the SM via Higgs-strahlung and WW fusion as a function of the center-of-mass energy.   

\begin{figure}[H]
   \centering
    \includegraphics[width=3in]{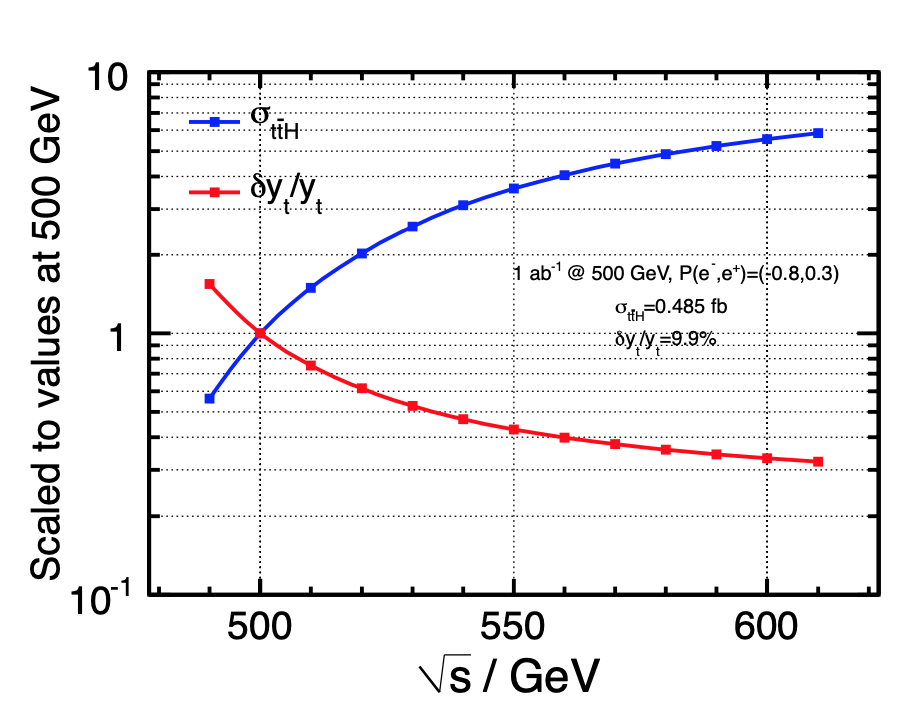}
    \caption{ttH cross section (blue) and uncertainties on top yukawa coupling (red) versus center of mass energy of the collider). Figure 19 from Electroweak Topical Group report.}
    \label{fig:top-yukawa-vs-cme.png}
\end{figure}

The direct measurement of the top-Yukawa coupling in \ee$\rightarrow ttH$ production requires a center-of-mass energy around 600 GeV (Figure \ref{fig:top-yukawa-vs-cme.png}). The cross section rises sharply around that energy; raising the center-of-mass energy to 550 GeV enhances the production rate by a factor or approximately four and the measurement of the ttH coupling by a factor two. Several groups have performed detailed full-simulation studies at center-of-mass energies ranging from 500 GeV to 1.4 TeV. With 4 ab$^{-1}$ at 550 GeV, a precision of 2.8\% is expected on the top Yukawa coupling, which could improve to 1\% with 8 ab$^{-1}$ at 1 TeV. Measurements at multiple center-of-mass energies and with different beam polarizations can further characterize the ttH coupling. 

Lepton colliders will also enable the search for new particles in the TeV range. Lepton colliders have a lower reach in energy compared to hadron colliders but excel in fully exploiting all possible manifestations of new physics within reach. 

\subsubsection{Low mass new particle searches}

There exists several well-motivated new physics scenarios which predict new low mass particles.  Often they are produced at low rates in decays of other heavy states, e.g., the H(125 GeV) state, and subsequently decay to SM particles. These particles may be long lived in some scenarios. Exploration of these scenarios is a focus of the LHC program presently. However, given large backgrounds at a hadron collider the low transverse momenta of the final state particles and long lifetimes of the new states is likely to leave a significant parameter space of the new theories left for exploration beyond the HL-LHC. At a trigger-less electron-positron Higgs factory with full recoil information and sensitive new detectors, the parameter space available for such new theories could further explored.

For instance, the exotic Higgs decay program is well studied summarized in Section 8.2 of the ILC~\cite{ILC} report to Snowmass 2021. Long-lived particle searchers were also studied in particular for the FCC-ee case~\cite{FCC-ee-LLP}.

\subsection{TeV-scale Machines (1 to 3 TeV)}

At an \ee-machine, ZH-process for producing the Higgs bosons dominates at sub-TeV energy scales. However, direct measurement of Higgs self-coupling requires significantly higher energies. At the TeV scale, the W-boson fusion process contributes significantly to di-higgs production.

Only linear \ee colliders are able to reach TeV scale. The collider cost driven by the accelerating gradient, power consumption and R\&D required to design a machine, sets the plane for discussion.

From physics point of view, the higher the energy reach the better given sufficient growth in luminosity. Looking at design maturity for machines at TeV-scale, there are TeV upgrade options for ILC, CLIC and C$^3$. Other options based on energy recovery Linacs, which require significant R\&D, are ReLiC, CERC ERLC, as well as wakefield accelerators.

The length of the linac is determined by the maximum operational acceleration gradient. The site geography and agreements with local authorities to grow in length are essential to scale up. Further, the design and length of the beam delivery system (BDS) is an important item that needs to be addressed from the very start, because a straight-forward upgrade of the BDS at a later  higher center-of-mass energy stage would be disruptive. Within, the constraints presented, the ILC at 1-TeV and CLIC at 3-TeV are suitable options to look at. Options at 3-TeV by C$^3$ and WFAs are also noteworthy.

One essential goal of a TeV-class \ee-machine is the measurement of the Higgs boson self-coupling using HH$\nu_e\bar{\nu_e}$ production in W-boson fusion channel. Study of the multi-vector-boson production at high scales allows study of electroweak symmetry breaking at unprecedented precision. Searches for electroweak produced new physics is also of interest, although at 1-3 TeV scale, the reach beyond the HL-LHC is somewhat limited.

\subsection{Energy frontier (10 TeV scale)}

The muon collider studies indicate that 10-TeV scale machine has a strong physics case. For an electron machine to reach that energy scale necessitates consideration of advanced technologies, both for acceleration and beam focusing. While there are no fully developed collider proposals at this time, there are promising technological directions that can be investigated further. The performance of the hadron machines the FCC-hh and the SppS are comparable to 10-TeV scale for electro-weak processes, whereas for the very highest mass colored objects, the 100-TeV hadron machines hold the advantage. Nevertheless, a 10-TeV \ee-machine, based on a muon or a WFA collider, if feasible, is an ideal follow on to a Higgs Factory.

A 10-TeV center of mass collider capable of producing ${\it O}$(10) ab$^{-1}$ data, will have essentially the same performance as the 100-TeV FCC-hh machine, when considering electroweak produced states with the advantage of having no pileup to mitigate. While the FCC-hh will have an advantage when it comes to searching for colored states and compositeness studies, the lepton colliders with such a reach have the benefit of excellent signal to background, taking advantage of the vector boson fusion production processes. Several physics cases for a 10-TeV muon collider with 10 ab$^{-1}$ were studied recently, and were shown to provide a compelling motivation. For instance the Higgs self-coupling measurement of few percent precision is possible using a 10-TeV, 10 ab$^{-1}$ data set. Feasibility study of a 10-TeV \ee~collider with the same luminosity profile, based on advanced acceleration and plasma lens system is of interest. To the first order we expect the physics case from the muon collider will carry over to the electron case. If anything, the absence of the large beam-induced-background and perhaps potentially larger angular coverage of the detector should help improve the measurements. Detailed studies are necessary to make any serious comparisons.

\subsubsection{Direct searches for high mass particles}

The Muon Collider Forum report~\cite{MuColForum} articulates that using a measured deviation of a Higgs Factory higgs-gluon coupling at 1\% from the SM expectation, one can obtain a constraint on stop mass, in degenerate stop-squark scenario, of $m_{\tilde t}$ of order a few TeV. On the other hand, the direct discovery reach for each stop squark at a high-energy lepton collider extends up to very nearly $m_{\tilde t} = \sqrt{s}/2$, or about 5 TeV for a 10 TeV center-of-mass collider~\cite{AlAli:2021let}. Thus, the 10-TeV collider will discover the physics responsible for the measured Higgs coupling deviation at the Higgs Factory. On the other hand, if the stops are sufficiently light, the measurement of the percent-level gluon coupling could play a role in elucidating the detailed structure of stop mixing with the extraction of the mixing parameter.

As another example, the Muon Collider Forum report~\cite{MuColForum} also makes the case, where in composite models, the Higgs couplings to $W$ and $Z$ bosons receive corrections of order $v^2/f^2$. This result follows from universal model-independent considerations when the Higgs is a pseudo-Nambu-Goldstone boson. For instance, in the minimal composite Higgs model, one obtains a bound on the scale of decay constant $f$. The decay constant $f$ does not directly determine the mass scale of all composite states. A naive dimensional analysis scale remains out of reach of a 10 TeV lepton collider if $f \sim 5\,\mathrm{TeV}$. However, composite Higgs models also contain other particles, like top partners at the scale $f \sim 5$ TeV, playing a major role in the naturalness of the theory. The precision measurement of the Higgs-W-boson coupling can  point to the parameter space where a direct discovery of new particles associated with compositeness can be made. Because this scenario involves tree-level states with strong coupling, it is one of the cases where precision is expected to be farthest ahead of direct reach, so this is an encouraging conclusion. 
	
A high energy muon collider 
can also make powerful statements about the electroweak WIMP Dark Matter for a fermionic DM particle in connection with its thermal relic abundance. In the universal and inclusive signals, the particles in an EW multiplet are produced in association with at least one energetic SM particle. The most obvious channel is the pair production of the EW multiplet associated with a photon, which dominates the sensitivity to higher-dimensional EW multiplets. Additionally, vector boson fusion (VBF) channels unique to a high-energy lepton collider \cite{Costantini:2020stv} can also contribute. In this case the mono-electron case should work like the mono-muon case studied for the muon collider. Availability of  high-energy lepton collisions, 10-TeV and above, can substantially improve the constraints on thermal dark matter, serving as one of the main physics drivers for a 10-TeV class energy frontier lepton colliders.

\subsubsection{Indirect probes of new physics scales}

Again adapting the muon collider forum report~\cite{MuColForum}, consider the Higgs oblique operator $\partial_\mu(H^\dagger H)\partial^\mu(H^\dagger H)$, which can affect the precision measurement of the di-Higgs production rate at a high energy lepton collider~\cite{Buttazzo:2020uzc}. One possible origin for such an operator is a singlet scalar mixing with the Higgs, as in the Twin Higgs scenario~\cite{Chacko:2005pe}. A 10 TeV lepton collider could probe $f \sim 10\,\mathrm{TeV}$ in direct searches for such a scalar $\phi \to hh \to (b{\bar b})(b{\bar b})$~\cite{AlAli:2021let}. In this case, the precision constraint and the direct search are extremely similar, with the former being a non-resonant search for the di-Higgs process and the latter a resonant search.

\subsection{Physics Timeline}

\begin{figure}[h]
    \centering
    \includegraphics[width=6in]{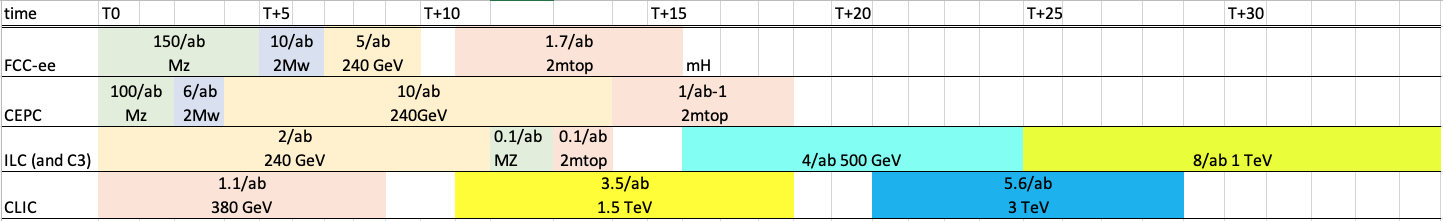}
    \caption{Physics timeline extracted from various run plans provided by the collaborations.}
    \label{fig:physics-timeline}
\end{figure}

Run plans used by the proponents of various collider options are compared in Figure \ref{fig:physics-timeline}.  While the run plans can be varied to optimize physics of interest this summary provides a visual representation of the time lines to obtain the results submitted to the Snowmass process in years and selected $\sqrt{s}$ choices, given the start time of the data taking ($T_0$). Circular colliders provide higher luminosities at lower center of mass energies and liner colliders have the possibility to extend the energy reach in \ee environment and provide polarization which can compensate in some cases the lower luminosities.

Table 1, 2 and 3 are a summary of the evaluation performed by the Integrated Task Force group that has been summarized in this report~\cite{ITF}. Table 1 shows the technical risk evaluation considering the technical risks of key components necessary for implementing the proposed facility, and design status of the facility. Table 2 shows the number of years that would be required by each accelerator to start the physics program assuming a technically driven schedule and ignoring geopolitical factors. Table 3 shows the estimated cost of each accelerator in 2021 B\$, including the excavation. 

\begin{table*}[ht!] 
    \centering
	\caption{\label{table:maturity-risk} Accelerator design status and risk based on the ITF classification \ee ~colliders: Design Status: I - TDR complete; II- CDR complete; III - substantial documentation; IV - limited documentation and parameter table; V - parameter table. Risk: The risk metrics takes into account the technical risks of key components necessary for implementing the proposed facility. I corresponds to lower risk and IV is higher risk.}
	\begin{tabular}{c | c  | c | c}
		{Collider} & {Type} & {Design Status} & {Risk}  \\
		\hline
		ILC (0.25 TeV) & Linear & I & I \\
		\hline
		FCC-ee (0.24 TeV) & Circular & II & I\\
        CEPC (0.24 TeV)    & Circular &  II & I \\
		CLIC (0.38 TeV) & Linear & II & I \\
        CLIC (3 TeV) & Linear &  II & I \\  
        \hline
		C$^3$ (0.25 TeV) & Linear & III & II  \\
		CERC (0.24 TeV) & Circular &  III & II\\
		\hline
		ReLiC (0.24 TeV) & Linear  &  IV  & II\\
		ERLC (0.24 TeV) & Linear & IV & II\\
		ILC (3 TeV) & Linear & IV & II\\
		C$^3$ (3 TeV) & Linear & IV & II\\  
		ReLiC (3 TeV) & Linear & IV & III\\
		WFA (3 TeV) & Linear & IV & III\\
		\hline
		WFA-flat (15 TeV) & Linear & V & IV\\
		WFA-round (15 TeV) & Linear & V & IV \\
	\end{tabular}\\
\end{table*}

\begin{table*}[ht!] 
    \centering
	\caption{\label{table:years-RD} Accelerator preproject RD on key technologies and years to first physics, taking into account the design maturity, years of pre-project RD, and risks, based on the ITF classification \ee ~colliders*}
	\begin{tabular}{c | c | c | c}
		{Collider} & {Type} & {years of preproject RD} & {years to first physics} \\
		\hline
		ILC (0.25 TeV)    & Linear   & 0-2  & $<12$\\
		\hline
		FCC-ee (0.24 TeV) & Circular & 0-2 & 13-18\\
       	CEPC (0.24 TeV)   & Circular & 0-2 & 13-18 \\
		CLIC (0.38 TeV)   & Linear   & 0-2 & 13-18 \\
        C$^3$ (0.25 TeV)  & Linear   & 3-5 & 13-18 \\
        \hline
        CLIC (3 TeV)      & Linear   & 3-5 & 19-24\\ 
		C$^3$ (3 TeV)     & Linear   & 3-5 & 19-24\\  
        ILC (3 TeV)       & Linear   & 5-10 & 19-24 \\
		CERC (0.24 TeV)   & Circular & 5-10 & 19-24\\
        \hline
		ReLiC (0.24 TeV)  & Linear   & 5-10 & $>25$  \\
		ERLC (0.24 TeV)   & Linear   & 5-10 & $>25$ \\
		ReLiC (3 TeV)     & Linear   & 5-10 & $>25$ \\
		\hline
		WFA (3 TeV)       & Linear   & $>10$ & $>25$ \\
		WFA-flat (15 TeV) & Linear   & $>10$ & $>25$\\
		WFA-round (15 TeV)& Linear   & $>10$ & $>25$  \\
	\end{tabular}\\
\end{table*}


\begin{table*}[ht!] 
    \centering
	\caption{\label{table:Cost} Accelerator cost based on the ITF classification \ee ~colliders}
	\begin{tabular}{c | c  | c}
		{Collider} & {Type} & {Cost [2021] B$\$$}  \\
		\hline
		ILC (0.25 TeV)    & Linear   & 7-12  \\
		CLIC (0.38 TeV)   & Linear   & 7-12  \\
		C$^3$ (0.25 TeV)  & Linear   & 7-12  \\
		ReLiC (0.24 TeV)  & Linear   & 7-12   \\		
		\hline
		FCC-ee (0.24 TeV) & Circular & 12-18 \\
       	CEPC (0.24 TeV)   & Circular & 12-18  \\
		ERLC (0.24 TeV)   & Linear   & 12-18 \\
		C$^3$ (3 TeV)     & Linear   & 12-18 \\  
		PWFA (3 TeV)      & Linear   & 12-18 \\
		WFA LC (3 TeV)    & Linear   & 12-18\\			
        \hline
        CERC (0.24 TeV)   & Circular & 18-30 \\
        CLIC (3 TeV)      & Linear   & 18-30 \\ 
        ILC (3 TeV)       & Linear   & 18-30 \\
        \hline
        LWFA (3 TeV)      & Linear   & 18-50 \\
        \hline
		ReLiC (3 TeV)     & Linear   & 30-50 \\
	\end{tabular}\\
\end{table*}

\section{Accelerator }\label{accel}

The accelerator community has developed a portfolio of \ee~technologies and concepts to meet the needs of the 3 energy sectors and has corresponding R\&D programs planned for the near-term (this Snowmass decade, 2025-35), midterm (this next Snowmass decade, 2035-45) and long-term (two Snowmass decades and beyond, $>$2055). \\

The needs of the \emph{Higgs and Electroweak Machines ($<$1~TeV)} can be at least partially met by all accelerators technologies (linear and circular) but the only option that is capable of beginning construction in the near-term is the ILC.  Midterm (2033-43) collider options at this lowest energy scale are available from both linear colliders, lead by CLIC-0.38TeV with a completed conceptual design report (CDR) followed closely by $C^3$-0.25TeV, as well as the circular colliders, FCCee-0.36TeV and CEPC-0.24TeV. Only linear accelerator technologies are extensible to \ee \emph{TeV-scale Machines (1 to 3 TeV)}, albeit with different challenges, footprints and performance. There are no near-term options at this energy scale but the ILC-1TeV is a midterm option.  Finally, reaching the \emph{Energy Frontier (10 TeV Scale)} will require extensive R\&D and is the focus of long-term accelerator R\&D under development by WFAs exploring how this could be feasible in terms of both reduced footprint and power consumption as well as as well as ERL-based accelerators, like ReLiC, ERLC and CERC, which aim to reduce power consumption (or increase luminosity).

Overall, we consider nine \ee ~accelerator projects in this section: six of these \ee~accelerator machines are \emph{linear}:  C$^3$, CLIC, ERLC, ILC, ReLiC and WFA while three are \emph{circular}: CEPC, CERC and FCCee. The Accelerator Section relies heavily on the \cite{ITF} report to generate our simplified summary of the \ee~collider projects but the interested reader can find more details in the ITF report~\cite{ITF}  and the various collider white papers.  We note that some machines like the SRF-based HELEN and circular ``site-filler'' Fermilab machines were not detailed in the ITF report~\cite{ITF} , and as such were not considered here\cite{FNAL-FC-Report}.

The primary goal of the Accelerator Section is to summarize the accelerator technology for each \ee ~collider project in order to assist the Snowmass community with decisions it faces, especially in near-term Snowmass decade \textemdash ~2025-35. All nine projects are summarized in~Table \ref{table:circular} and Table \ref{table:linear}. These tables summarize each project in terms of their key accelerator cost drivers (Power Consumption and Length), Design Status and their near-term R\&D targets~\cite{ITF}. \emph{Design Status} is taken directly from the ITF report's Table 14 and it indicates the current status of the design concept: I - TDR complete, II - CDR complete, III - substantial documentation, IV - limited documentation and parameter table, and V - parameter table. High design status  accelerators are less mature which implies they need more and longer-term R\&D to realize the potential of their respective technologies. The last column reflects the priority R\&D that needs to be done in the near-term . 

\begin{table*}[ht!] 
    \centering
	\caption{\label{table:circular} Core Accelerator Parameters of circular \ee ~colliders (R $\approx$ 100~km), taken from the ITF report}
	\begin{tabular}{c | c | c | c}
		\rowcolor{gray}
		\textcolor{white}{Collider} & \textcolor{white}{Power Consumption (MW)} &  \textcolor{white}{Design Status} & \textcolor{white}{near-term R\&D}\\
		\hline
		FCC-ee (0.24 TeV) & $280$ & II & RF power, Positron, Magnets\\
		CEPC (0.24 TeV) & $340$ &  II & RF power, Positron, Magnets\\
		CERC (0.24 TeV) & $90$ &  III & High Energy ERL\\
	\end{tabular}\\
\end{table*}
\begin{table*}[ht!] 
    \centering
	\caption{\label{table:linear} Core Accelerator Parameters of linear \ee ~colliders, taken from the ITF report}
	\begin{tabular}{c | c | c | c | c}
	    \rowcolor{gray}
		\textcolor{white}{Collider} & \textcolor{white}{Power  (MW)} & \textcolor{white}{Length (km)} & \textcolor{white}{Design Status} & \textcolor{white}{Near-Term R\&D} \\
		\hline
		ILC (0.25 TeV) & $140$ & $14$ & I & \textemdash\\
		C$^3$ (0.25 TeV) & $150$ & $3.7$ & III & demonstration facility\\
		CLIC (0.38 TeV) & $170$ & $13.4$ & II & TDR completion\\
		ReLiC (0.24 TeV) & $100$ & $20$ &  IV & kickers \& damping rings\\
		ERLC (0.24 TeV) & $250$ & $60$ & IV & RF power\\
		\hline
		ILC (3 TeV) & $400$ & $59$ & IV & TW SRF structures\\
		C$^3$ (3 TeV) & $700$ & $26.8$ & IV & HOM damping and IP stability\\ 
		CLIC (3 TeV) & $580$ & $42$ & II & two beam acceleration\\ 
		ReLiC (3 TeV) & $450$ & $360$ & IV & kickers and damping rings\\
		WFA (3 TeV) & $170-340$ & $0.44-20$ & IV & Integrated Design Study\\
		\hline
		WFA-flat (15 TeV) & $1121$ & $10-50$ & V & Integrated Design Study\\
		WFA-round (15 TeV) & $90-210$ & $10-50$ & V & Integrated Design Study\\
	\end{tabular}\\
\end{table*}
\subsection{Higgs and Electroweak Accelerators: ($\le$ 1 TeV)}

\underline{Near-term.} In the scenario where the decision is made to begin construction of a collider in the $\le$ 1 TeV energy regime before the next Snowmass process (i.e. within the first Snowmass decade after P5), then there is only one \textbf{near-term option} that is considered ``shovel ready'': the ILC-0.25TeV. It is the only \ee ~collider that is \emph{TDR complete} (Table \ref{table:linear}) and was the only collider concept in the ITF report to have its technical design status ranked 'I', to be given its highest rating, $I$, in all categories shown in ITF Tables 14 and 15, and was judged to have the shortest ``Time to First Physics'' as shown in Table 17 among \textbf{all} $24$ colliders considered: lepton, hadron or $ep$.  We recognize that the construction of a collider may not be given soon, likely due to lack of affordability rather than the lack of an interesting physics,

While we await for the decision to build an \ee ~collider, the HEP community must carryout out near-term, critical accelerator R\&D, develop of CDRs and TDRs, and identify synergies between the various colliders so that \ee ~collider options. In the midterm and long-term, the US HEP community should broadly support accelerator research in order to keep all roads open to mitigate the substantial uncertainty in funding a near-term collider at the current cost levels.

\underline{Midterm.}  The midterm is defined by the next Snowmass decade of 2032-42 (approximately equivalent to the combined ITF time scales of 13-18 and 19-24 years).  The leading midterm collider option at this energy scale is the linear collider CLIC@0.38TeV with its completed CDR. Overall, midterm options include two linear collider options, $C^3$-0.25TeV and CLIC-0.38TeV, and two circular collider options, CEPC-0.24TeV and FCCee-0.24TeV.  Each of these midterm options require near-term accelerator R\&D to advance the critical technology elements as identified in Table 7 of the ITF report and listed in Table \ref{table:circular} and Table \ref{table:linear}. The CLIC-0.38 TeV project requires support to move from the CDR level to the TDR level.  The C$^3$-0.25TeV project is targeting a demonstration facility~\cite{CCC-Snowmass} to perform a string test of 3, 9-meter cryomodules assembled in series and powered by 18 RF sources. Both ERL concepts require High-Q, low-frequency SRF cavities. On the circular side, the needs of FCCee and CEPC include high-power RF cavity and high-power sources to accommodate synchrotron radiation losses, positron sources and magnet development.  

\underline{Long-term.} If no collider is built in the near-term or mid-term, then the longer-term projects begin come into play. The projects need support to carryout research now.  These include the ERL-based concept, ReLiC-0.25TeV, ERLC-0.25TeV, and CERC@0.6TeV. ReLiC requires high rep-rate kickers.

\subsection{TeV-scale (1 to 3 TeV) accelerators}

There is no near-term option to begin construction of an \ee collider in the 1-3 TeV energy regime, as no accelerator concept has currently reached the required level of maturity (e.g. a TDR).  Further, only linear colliders are able to reach this energy scale so no discussion of circular colliders is given in this or the next subsection. The collider options at this energy scale are all midterm and long-term. 

The midterm options at the 1-3 TeV energy scale have similarities with the midterm options for the $\le$1 TeV energy scale.  The most mature linear collider concepts at this energy scale is the CLIC-3TeV, with a completed CDR and risk status of I as shown in Table \ref{table:linear}. The CLIC-3 TeV option is followed closely by $C^3$.  The CLIC-3TeV project requires support to move from the CDR level to the TDR level but, in addition, the midterm roadmap requires further development of its two beam acceleration method.  The $C^3$-2TeV project midterm R\&D focuses on incorporating HOM damping and demonstrating stability at the IP. The midterm ReLiC roadmap continues to require high rep-rate kickers but also HOM detuning. ERLC

The long term options at the 1-3 TeV energy scale are focused in ways to dramatically reduce cost by developing novel acceleration methods. WFA's have great potential to reduce the facility footprint and increase accelerator efficiency while ERL-based colliders aim to reduce power consumption (or increase luminosity).  The facility footprint drives the civil engineering cost and is mostly due to the length of the linac. Therefore, one of the primary goals of WFA program is to dramatically increase the operational acceleration gradient to reduce the linac length.  In addition to construction cost, both WFAs and ERLs seek ways to reduce site power. WFAs look for way to increase the rf/laser power to beam efficiency and ERLs try to recover the spent beam energy.

\subsection{Energy frontier (10 TeV scale) accelerators}
There are no near-term or midterm accelerator options at this energy scale.  Accelerator technologies are striving for a path to enabling  \ee ~colliders at the $O(10)$-TeV scale while providing $O(10)$ ab$^{-1}$ within a manageable footprint. I.e.  all options here require long-term R\&D.  The difficulties in this energy regime are two-fold: cost and the physics at the Interaction Region (IR).  Regarding cost, long-term accelerator options require the same R\&D program as defined in the long-term section for the 1-3 TeV machines.  In addition, \ee colliders at this highest energy scale require R\&D regarding the physics at the IR.  Traditionally, \ee ~colliders were considered to have an upper limit of  3-TeV due to beamstrahlung effects at the IR, but advanced acceleration concepts are under development in the WFA community to extend \ee~colliders into this regime by exploring very short bunch lengths to  mitigate beamstrahlung. In addition, plasma lens based beam delivery systems need further investigation. Continued US involvement in the development of advanced accelerator technologies is in the interest of the global community.


\section{Detectors }\label{detect}

\subsection{Detector Requirements}

Detectors for all $e^+e^-$ colliders must fulfill similar performance requirements driven by the physics goals \cite{ILC, CLIC-CDR, FCC-Snowmass}. 

The detectors have to precisely reconstruct the Higgs recoil, Higgs decays to muons, and potentially new physics signals involving leptons. 
This requires efficient and precise tracking with momentum resolutions of $\sigma(p_T)/p_T\ \approx\ 0.2\%$ for $p_T<100$ GeV and $\sigma(p_T)/p_T^2\ \approx\ 2\times10^{-5}/\rm{GeV}$ for $p_T>100$ GeV. 
For an all-silicon tracker, this corresponds to about 2\%\ $X_0$ and a 7 $\mu \rm{m}$ point resolution per measurement layer. 
The impact parameter resolution is driven by flavor tagging requirements to be $\sigma(d_0)\ \approx\ (5\oplus(10 - 15)/(p\sin^{3/2}\theta))\ \mu \rm{m}$ which requires $< 0.2\%\ X_0$ and about 3 $\mu \rm{m}$ point resolution per measurement layer in the inner part of the tracker.

The calorimeter performance is driven by the need to distinguish hadronic decays of W and Z bosons. This leads to a requirement for the jet energy resolution of $\sigma(E)/E\ \approx\ $3-5\% for $50\ \rm{GeV}<E<100\ \rm{GeV}$.

The detectors have to be able to identify electrons and muons up to the highest energies and to tag jets with the flavor of their primary quark. Precise timing information is useful for particle id ($\approx$ 10 ps) and to reject out-of-time backgrounds ($\approx$ 1 ns).

A typical $e^+e^-$ collider detector has the following main features: 
\begin{itemize}
    \item An inner tracker with the lowest possible mass and the smallest possible inner radius.
    \item A low-mass outer tracker for excellent momentum resolution at high energies.
    \item A highly segmented calorimeter.
    \item A large superconducting solenoid. 
    \item A muon system
    \item Triggerless readout.
\end{itemize}

Detectors for the ILC have been studied in great detail to the level of a conceptual design report (ILD~\cite{ILD} and SiD~\cite{SiD}). The CLIC collaboration has adapted the ILC detectors to the requirements at higher energy~\cite{CLIC-CDR, CLIC-DetectorConcept}. For circular colliders, the CLD detector~\cite{CLD} for FCC-ee and the baseline detector for CEPC~\cite{CEPC-CDR} were adapted from the CLIC design and the IDEA detector~\cite{IDEA} was proposed as a new concept. 

There are some differences in the environment at the different colliders which affect the detector requirements. 

For linear colliders the beam is divided into trains of closely spaced bunches, separated by longer gaps (ILC $\approx200$ ms\cite{ILC-Snowmass}, CLIC $\approx20$ ms\cite{CLIC-CDR}, CCC $\approx 8$ ms\cite{CCC-Snowmass}). These gaps allow for power pulsing of the readout electronics at linear colliders. The electronics is only powered during the bunch trains and powered off during the long gaps. This reduces or eliminates the need for active cooling systems which are a dominant source of dead mass for tracking detectors. This enables sufficiently low mass trackers to achieve the required momentum and impact parameter resolutions. 
In many cases the calorimeter is fit inside the magnet in order to minimize the dead material before the calorimeter.

A strong magnetic field ($>$3T) is desirable to curl up the soft $e^+e^-$ pair background from beamstrahlung. The inner radius of the tracker is limited to about 3 cm by this $e^+e^-$ pair background. 

Pileup of hadronic two-photon interactions increases from 0.3/bunch crossing at 500 GeV to 3.2/bunch crossing at 3 TeV. At 3 TeV this deposits 20 TeV of energy in the detector over a 156 ns bunch train. 
Timing measurements with a resolution of order ns are needed to reduce beam induced backgrounds. The expected fluence in the inner tracker is approximately $6\times10^{10}/\rm{cm}^2/\rm{year}$ and the total dose is approximately 300 Gy/year.

At circular $e^+e^-$ colliders, running at the Z pole with maximum luminosity imposes the strongest requirements on the detector. The magnetic field of the solenoid is limited to around 2 T in order to preserve the low emittance of the beams~\cite{arxiv:1905.03528}. In order to preserve the momentum resolution this necessitates a larger radius for the tracking system. Circular accelerators also have a high beam crossing rate which makes power pulsing impossible. The resulting need for an active cooling system makes achieving the required material budget in the tracker much harder. More R\&D is needed to reduce the mass in the tracker by using monolithic detectors. Requirements at the Z peak include that the acceptances must be precisely known which implies that detector boundaries have to be known to microns. The DAQ system has to deal with a rather high event rate of almost 100 kHz.

While detectors for the $\sqrt{s}\ <\ 1$ TeV case have been thought through to the level of conceptual design reports  and successfully adapted to the 1-3 TeV region (CLIC\_ILD and CLIC\_SiD), there has been much less formal optimization for detectors at $\sqrt{s}\ \sim\  10$ TeV. Clearly the calorimeters will need to be deeper to contain the showers, the trackers will need to be longer and/or have higher point resolution to measure the highest $p_T$ tracks, and the magnetic fields will need to be higher (for the same reason). While initial designs of detectors for a 10 TeV machine will be based on the technologies we mention above and below, we should expect dedicated R\&D will be needed to solve problems particular to these detectors. The investigation of large solenoids with fields of $>~5$ T is one such example.

\subsection{Particle Flow}

The requirement to separate W and Z bosons in their dominant hadronic decays translates to a mass resolution of about 2.7\% which implies a dramatic improvement in jet energy resolution compared to current collider detectors. 
Typically, calorimeters are divided into em and hadronic sections with the hadronic section often coarsely segmented such that many particles from a jet end up in the same segment. This means that the resolution is not given by the intrinsic precision with which the calorimeter measures energy deposited but is dominated by the fluctuations in the fragmentation of jets which consist of charged and neutral hadrons. Neutral pions decay to photons adding em showers into the hadronic calorimeter. The response of a calorimeter differs for these and if the segmentation is not sufficient to separate them only an average response can be applied. 

The particle flow technique~\cite{arxiv:2203.15138} is based on highly segmented calorimeters that allow to continue to track the path of a particle into the calorimeter and assign calorimeter signals to specific particles. Together with the tracker information neutral and charged hadrons, electrons, and photons can be identified, the calorimeter response can be corrected appropriately, and tracking information can be included into the reconstruction of each particle as well. Jet energies can then be reconstructed from the sum of the particle energies rather than from calorimeter signals corrected only for an average response, making resolutions required to distinguish hadronic W and Z decays possible.

\subsection{Tracking Detectors}

The primary components of most collider tracking systems are inner and outer trackers. Outer trackers are more likely drift chambers,  including Time Projection Chambers (TPCs) - there are many examples of these over the past 50 years - or silicon (strip) trackers. The current example of a silicon strip outer tracker is CMS (not an $e^+e^-$ detector) and the  most developed proposed $e^+e^-$ example is SiD. Drift chambers are preferred for less material in the tracking volume and better dE/dx measurements (especially TPCs). Strip trackers have more (support, cooling and sensor) material in the tracking volume but better point resolution, two-hit separation, and less pile-up of background.

Drift chambers are also perceived as less-expensive than strip trackers, and that may be of use when colliders are just turned on (likely at lower energies) after an expensive construction project, and where less multiple scattering is desired and cleaner events may be available. Another important function of the outer tracker is projection to the face of the calorimeter, and here strip trackers are generally better, though several designs have one layer of Si outside a drift chamber to have one last high precision point on a track for momentum resolution and track matching, and to use as a timing detector (TOF).

The outer trackers seem to have a $\sqrt{s}$ selection - the lower mass drift chambers and TPCs, with dE/dx and lower costs, would seem to be a better match at low $\sqrt{s}$, and the strip trackers would seem more necessary at high $\sqrt{s}$ with their better spatial resolution.  

Inner trackers, on the other hand, would seem to have a closer mapping to machine type. The type of inner tracker used at SLD (very thin CCDs) works exquisitely with the \ee\ environment, but requires a long gap of no beam to read out the hits (from many bunch collisions) without beam-induced noise. While this sort of detector might seem optimal for linear colliders, the bunch-train structure for even these varies from machine to machine. The main advantage of linear collider bunch structure for tracking detectors now seems to be the relaxed cooling (and therefore material) budget that comes with using pulsed power. The leading candidate for this type of detector now seems to be MAPS~\cite{arxiv:2203.07626}. These seem to be useful for inner trackers, outer trackers, and calorimeters (all with different optimizations).    
Another leading candidate for the the inner tracker sensors are very thin hybrid sensors (sensors bump-or-otherwise bonded to electronics). Both of these inner tracker options are being pursued for the LH-LHC upgrades - thin MAPS detectors for the ALICE inner tracker and thin (but not thin to the extent needed for $e^+e^-$ detectors) hybrid detectors for ATLAS and CMS. However, R\&D will still be needed for \ee\ requirements in this area over a similar timescale.

\subsection{Calorimeters} 

The main characteristic of a particle flow calorimeter is fine granularity and consequentially large channel count. The CALICE collaboration is dedicated to the development of a particle flow calorimeter. The ILD, SiD detectors for ILC, the CLICdet detector for CLIC, and CLD detector for FCC-ee, and the baseline detector for CEPC all use similar technology concepts. Typically, in the em section, scintillator strips with SiPM readout and silicon diodes are options for the active material with Tungsten as passive absorber. Typically the granularity is 5 mm $\times$ 5 mm. For the hadronic section scintillator tiles with SiPM readout and RPCs are options for the active material with steel as absorber with a typical granularity of 3 cm $\times$ 3 cm. Conventional analog readout as well as digital readout systems are being considered. 

Prototypes of various designs have been built and beam tested. R\&D needs to be carried out for various aspects, depending on technology. A common focus of R\&D is to scale these prototype designs to larger detectors and to study their performance with realistic particle showers. Given the large number of channel counts, scalable production and assembly techniques are critical to reduce manual steps.

Dual readout calorimeters provide another way of overcoming the limitation of the jet energy resolution by the different responses to em and hadronic showers~\cite{arxiv:2203.04312}. The RD-52/DREAM collaboration developed the technology for a calorimeter that reads out both scintillation light produced primarily by hadrons and Cherenkov light emitted mostly by relativistic electrons.

Another topic of R\&D is a compact readout system to service the large number of readout channels. Since it is not possible to bring the large number of analog signals out of the detector the front-end electronics has to embedded into the active layers, requiring high reliability and radiation hardness. In order to achieve optimal resolution, calibration also is an important consideration. 

Time measurements with ns resolution for calorimeter signals are useful to reject beam-gas background in $e^+e^-$ colliders and are being considered for these calorimeters.

\subsection{Particle ID} 

Particle ID was an important part of the LEP and SLC programs at the $Z$ (and above at LEP), and will continue to be useful at Higgs factories, above \tt\ threshold, and to the highest $\sqrt{s}$.

The primary means of identifying particles containing $b$ and $c$ quarks is through their decay (vertex) properties measured in the inner tracker. This is what drives the precise $d_0$ measurement requirements. Lepton identification (electrons from calorimeters and muons from instrumented magnet flux return) is also important for tagging heavy flavor.

TPCs are best for measuring dE/dx. Drift chambers are not as good and the information is not as frequently used as with TPCs. Silicon trackers only provide some dE/dx discrimination in limited momentum ranges. Algorithms to improve dE/dx measurements are also important and are constantly being studied in detector physics.

Projection of tracks to the calorimeter face (for improved $e/\gamma$ overlap reconstruction and less track/shower matching confusion) is usually  better with a silicon tracker than a drift chamber, but this can be remedied with a silicon ``wrapper'' (one layer of silicon detectors, for example) outside the drift chamber. If the right system is used (LGAD detectors, for example), this layer can also provide some time-of-flight information for the various tracks which in some cases can be useful for particle ID, and so is useful even with silicon trackers.

Modern versions of Ring Imaging Cherenkov (RICH) detectors used by DELPHI and SLD are also being studied~\cite{arxiv:2203.07535}, and other PID detectors of this ilk (transition radiation detectors, anyone?) should be also. 

\subsection{R\&D Needs}

R\&D needs for high energy physics were outlined in the DOE Basic Research Needs (BRN) study~\cite{BRN-HEPinstrumentation}. 
The R\&D program for $e^+e^-$ collider detectors has been going on for many years. Nevertheless R\&D must continue to incorporate new technologies into the existing detector designs. New and emerging technologies should be investigated to improve detector performance and/or make its construction more cost effective. The further reduction of the material in the trackers to minimize multiple scattering is an continuing goal of R\&D. Finally, a significant engineering effort is required to design support structures and services to turn prototypes into actual full-size detectors. This section outlines some of the main R\&D areas \cite{ILC-Snowmass, FCC-Snowmass}.

\begin{itemize}
\item Monolithic Active Pixel Sensors (MAPS) implemented with the rapidly developing CMOS technology~\cite{arxiv:2203.07626} may provide the technology to achieve some of these goals by pushing functionality down into the sensor and further increasing the granularity of sensors. They are being explored for both trackers and em calorimeters. Larger, multiretical sensors may be possible to reduce the amount of material needed for the support structure.

\item Low-Gain Avalanche Diodes (LGAD) provide the potential to reduce the thickness of sensors and to perform precise time measurements. R\&D is needed to optimize the sensor design and to adapt this technology for use in a large precision tracker system~\cite{arxiv:2204.00149}. 

\item Using 3D integration may reduce the material in pixelated sensors-readout assemblies and improve their performance and reliability~\cite{arxiv:2203.06093}. 

\item Thin film detectors have the potential to drastically reduce the amount of material in tracking detectors and to slash their cost by orders of magnitude~\cite{arxiv:2202.11828}.

\item Submicron pixels with a quantum well can improve the position resolution of vertex detectors to 0.5 $\mu$m~\cite{arxiv:2202.11828}.

\item Pushing the precision of the timing of signals in the tracker and the calorimeter into the 10 ps range would significantly aid particle id. In the presence of larger backgrounds, e.g. at a 3~TeV linear collider, precision timing can help reduce these backgrounds~\cite{arxiv:2203.07286, arXiv:2005.05221}.

\item R\&D is also required to develop technologies to efficiently distribute power and cooling to these highly granular detectors. Serial powering and microchannel cooling are being investigated.

\item An important area of R\&D is to make the technologies scalable to enable the construction of a large detector. This was outlined in the DOE BRN grand challenge to use integration to enable scalability for HEP sensors. 
\end{itemize}
In order to address these R\&D questions and design and develop the detector concepts further, a strong R\&D program is needed. This program should support R\&D work that goes beyond generic R\&D to support the specific challenges posed by the detector required for $e^+e^-$ colliders. Such a program needs to start now for the technology to build a full scale $e^+e^-$ collider detector to be ready when the HL-LHC program is completed. It can be seen from the HL-LHC detectors that the time between CD-0 (2016) to completion of the actual detectors (expected in 2027) can easily span over a decade. 

Besides funds, R\&D also requires the available of individuals with the required expertise to carry out the work. It is especially important to provide career paths for technicians and engineers in order to maintain this expertise in the field.


\section{Synergies }\label{syng}

Many examples exist of work or discoveries in one field of Physics or HEP helping another. Listed below are synergies for a few possibilities for a new \ee\ collider and associated detector.

\subsection{Accelerator Synergy}
The central challenge for HEP is reducing the cost of a future collider, and finding synergies with other accelerator users is a key way to ease burden to the HEP budget.  Electron accelerators are ubiquitous instruments employed thought-out society and makeup the vast majority of the nearly $\sim 30,000$ accelerators in operation across the globe today.  Applications span from mature industrial and medical uses to the cutting-edge R\&D for ultimate storage rings and linear accelerator-based X-ray free electron laser (XFEL).  Among the variety of accelerators under consideration to serve as the next collider, electron accelerators are uniquely positioned to leverage these global efforts thus opening a viable path to a linear collider.

Beam physics synergies are strong between all of the \ee linear colliders and XFELs.  Luminosity benefits from brighter electron beams through reductions in transverse emittance while shorter bunch lengths help to mitigate beamstrahlung.   Fortunately, accelerator research for XFELs at large user facilities like LCLS-II and European XFEL are also pursuing bright electron beams with symmetric transverse emittances of 0.1 um and bunch length towards 10s of fs. The short bunch immediately benefits the linear collider while a flat beam transformer would be used to generate asymmetric emittance required.

Accelerator technology synergies are specific to the \ee linear collider and include: the development of superconducting rf technology, EuXFEL and LCLS-II, to support the ILC main linac and the driver technology for the WFA schemes, SWFA and PWFA. A third example is the synergy between the proposed Compact Light facility and CLIC.  Wakefield accelerators have many synergies with the international community such as the recently funded EuPraxia facility, LaserNET USA and the many laser plasma accelerators. WFAs have been used to drive free electron lasers, as recently demonstrated by three experiments in Europe and Asia.

\subsection{Detector Synergy}

Some new technologies envisioned for $e^+e^-$ collider detectors can be validated at current and near term detector construction projects, such as the LHC phase 2 upgrades, experiments at EIC, or smaller experiments such as Mu3e.

Following on the previous type of synergy, it might be possible to imagine certain components of a detector, for example the superconducting magnet and the return flux might be optimized for both an $e^+e^-$ detector and later a $pp$ or $\mu^+\mu^-$ experiment, with the possibility to install new muon planes in the return flux in between. While this occasionally happens now (the BaBar solenoid is now the sPHENIX solenoid), it mostly happens without advanced planning and therefore without joint optimization.

At $e^+e^-$ colliders, the particle flow technique serves to separate the information of individual particle from each other to obtain optimal resolution. The technique can also be employed at Hadron colliders where the separation of individual particles allows the per particle removal of pileup effects from the hard scatter events. This has already been implemented by the CMS Collaboration in the current detector\cite{CMSParticleFlow}. The CMS phase 2 endcap calorimeter with its high granularity will be able to exploit the benefits of the particle flow technique even more. It should be capable to distinguish quark and gluon jets, for example\cite{CMSHGCal}.



\section{Path forward }\label{path}

We are entering into a period of great opportunity in High Energy Physics for future colliders. While the run plans for High-Lumi LHC will carry us through the next decade, the foundation for what comes next will be built this decade. An \ee~Higgs factory with 1\%-level measurements of Higgs boson couplings is widely recognized as the next critical machine.  An \ee~Higgs factory program will compliment the physics of HL-LHC and deliver the needed precision for the next stage of scientific discovery. An important consideration when choosing which machine to build next will be the ability to pursue energy upgrades if warranted. While the initial Higgs physics program should measure Higgs couplings with high accuracy and make an initial measurement of the Higgs self coupling, an energy upgrade into the 1-3 TeV range will greatly improve the Higgs self-coupling measurement and  allow unprecedented precision in the study of electroweak symmetry breaking. Ultimately, the \ee ~physics program may point to new physics at the O(10~TeV) scale, rendering the possibility of further energy upgrades of great interest. The concepts for future \ee~colliders that could fulfill this program vary widely from facility scale to technical maturity, and provide an exciting set of alternatives for the community to consider. 

Serious consideration is being afforded to FCC-ee and CEPC with feasibility studies underway to understand the scale of effort needed to construct these facilities with O(100km) circumference. These circular collider concepts would naturally follow with the operation of hadronic machines with a center of mass energy reach of O(100 TeV). Machine studies and technology pilot programs are proceeding over the next decade to address needed gaps in the maturity of these designs.

The path forward for \ee linear accelerator technology depends on the development of the optimal way to move from the weak scale, through the TeV scale and onto the O(10) TeV scale.  There are many possibilities under consideration but the overarching goals of high geometric gradient and low power consumption become more demanding as the energy scale increases.

The ILC remains the most technically advanced \ee- Higgs factory under consideration, closely followed by CLIC for linear collider facilities. Both of these have run plans that bring them into the range of TeV-scale machines with extensions of the main linac. ILC and CLIC have significant efforts to expand the adoption of relevant accelerator technology in other facilities and applications to increase their technical maturity. For superconducting rf technology, EuXFEL and LCLS-II can both be considered as string tests for the ILC main linac. The proposed Compact Light and EuPraxia facilities would serve a similar role by utilizing CLIC technology and wakefield technology for their accelerators.

A new concept, C$^3$, is also proposed for a compact Higgs Factory with energy reach of 550~GeV in an 8~km footprint. C$^3$ utilizes cryogenic-copper to achieve high accelerating gradient efficiently. A demonstration program over the next decade for C$^3$ could bring cryogenic-copper accelerator technology into consideration in the event a decision is not reached on more mature technologies. 

While these facility concepts are being evaluated for construction, accelerator technology continues to advance. Novel accelerator technologies are being pursued to enable more compact facilities and increase the energy reach of colliders. In a strong synergy with other linear collider concepts, wakefield accelerators will focus on integrated design studies to understand their potential for providing greenfield multi-TeV \ee~facilities as well as upgrades to other linear collider concepts to leverage off the infrastructure of a Higgs factory. This WFA integrated design study will be informed by the ongoing experimental programs at FACET-II, BELLA and AWA that are investigating R\&D challenges for WFA. Increases in luminosity and overall efficiency would also be highly desirable, and energy recovery concepts for both linear and circular machines could open up new frontiers in precision with high efficient machine concepts. Facilities, such as CBETA \cite{PhysRevLett.125.044803}, will provide valuable guidance on the viability of this approach. SRF technology will also continue to advance with material and cavity design studies (e.g. traveling wave SRF) aimed at increasing accelerating gradient, efficiency and luminosity.

Advances in technology will open new pathways for enhancing or upgrading linear collider facilities to reach higher energies. Options exist for upgrading gradient by replacing portions of the main linac with higher gradient technology or increasing the length of the site, both of which will require serious consideration of cost and power requirements. Early-stage concepts for ILC/CLIC/C$^3$ are all evaluating higher gradient structures and increased length as options to reach into the few TeV range. One extremely appealing scenario is that either after a Higgs factory or a $\approx$TeV upgrade, WFA technology reutilizes the existing site infrastructure as appropriate (\textit{e.g.} tunnel, damping rings, power distribution) with GeV/m gradients to reach up to the O(10 TeV) scale, thus providing a rich multi-decade physics program.

Many of the accelerator and detector concepts that are being considered share common research and development needs allowing for synergistic efforts that can be pursued over the next decade. Commonality is evident in the electron sources, positron sources, rf sources, beam delivery and beam dynamics challenges that are faced by these facilities.  Strong and targeted investments in accelerator research could yield significant breakthrough to help facilitate the realization of an \ee ~collider.

Given the excitement in the physics community, the viability of a future collider and the plethora of options, the time is also fast approaching to to pursue the R\&D needed to address the specific challenges posed by the detector required for an $e^+e^-$ collider. Pursuing \ee ~accelerator and detector R\&D vigorously is the best path for maximizing the physics potential of the next frontier machine and produce timely exciting results. 

\section*{Appendix: The \ee-Forum Mandate}

   The international community is making great progress in the development of a strategy
 for e+e- colliders as the next major projects after the end of the LHC program. To capture the growing interest for such colliders in the US community and further foster such studies in Snowmass, we propose to form an “e+e- Collider Forum” led by EF, IF and
 AF.

The intent of this Forum is to initiate dialogue and discuss differences and complementarities
between the various  e+e- collider concepts, either linear or circular, from the accelerator, detector and physics perspectives, in harmony with the rest of the wider international community.  The discussions
 will be informed by the Agora on future collider events, contributed white paper,  and works carried on within the topical groups. The outcome of these discussions will be summarized in a report that will
 serve as input to the Snowmass frontier reports. The Forum however will not replace or duplicate the
work that is carried out in the various topical groups that oversee all physics studies and are ultimately responsible for the final reports. Needless to say that the Forum’s activities will be carried out
 in close collaboration with the TG convenors of the relevant frontiers.

\section*{Appendix: Energy Frontier Q\&A} 

\begin{itemize}
    \item What physics is lost when we choose a circular collider and vice versa?
    
    Circular \ee colliders can provide very high luminosities to perform precision measurements but have limitations in the center of mass energy reach due to synchrotron radiation. FCC-ee and CepC have a full program of electroweak  measurements, as well as Higgs and top precision measurements based on running scenarios from the Z pole up to energies above the tt threshold production with very large statistics. The full program is expected to be completed in 15 years for FCC-ee and around 20 years for CEPC. FCC-ee and CEPC will not reach $>$ 500 GeV regime, required to access multi-Higgs boson production processes, mainly ZHH and HH$\nu_e\bar{\nu_e}$, which depend directly on the Higgs boson self-couplings, and the ttH production which directly measure the Higgs boson coupling to the top quark. 
    
    Linear colliders (ILC and $C^3$) are lower in cost, require less power and are more compact. Compared to circular colliders they do have lower luminosity, needing significantly longer running time to achieve the same level of precision for measurements at lower center of mass energy, despite of the polarization. For comparison at $\sqrt{s}=250$, FCC-ee  will collect 5ab$^{-1}$ in three years, CEPC 10ab$^{-1}$ in ten years and ILC 2ab$^{-1}$ in eleven years of operation, as summarized in Table 1. The circular collider luminosity advantage at the Z-pole is even larger. Although, the linear collider Z-pole electroweak measurements are likely less significant, in spite of the polarization advantage that they hold, the Higgs couplings measurements are quite comparable, when including the 500-GeV operation, with the additional benefit of direct di-Higgs access. ILC (one could extrapolate $C^3$), requires more than 30 years from the start of operation to collect the ZHH and HH$\nu_e\bar{\nu_e}$ samples (4ab$^{-1}$ at $\sqrt{s}=500$GeV and 8 ab$^{-1}$ at $\sqrt{s}=1$TeV) to reach 10\% precision in the measurement of the triple Higgs coupling~\cite{ILC-Snowmass}.
    
    Based on the Integration Task Force (ITF) report we have summarized in Table 2 the different e+e- collider proposals submitted to Snowmass.   
    
    Other e+e- colliders proposed using energy recovery Linacs, or Plasma wake field acceleration promise to reach high center of mass energies and very high luminosities, but require significant R\&D that would push the start of operation more than 25 years. 
    
\item What physics is delayed [i.e. will take a much longer time scale], even if they can be studies by both collider?
	
Here again we are confronted with the time it would take to have the accelerator in operation and the time it takes to have the integrated luminosity required to perform precision measurements. FCC-ee will take longer to start operations, given the level of maturity, compared to ILC, but it will collect 5ab$^{-1}$ at 250GeV in three years, while ILC needs 11 years to collect 2ab$^{-1}$. Polarization is an advantage of linear colliders (CERC - ciruclar can also provide polarization). The upgrade of ILC is to increased energy, up to 500 GeV, then 1 TeV, while the upgrade of FCC-ee will be a hadron collider at 100 TeV. 
	
	The primary consideration for the timely delivery of physics results is the start time of the physics program. Given the existence of a TDR and established site, the ILC holds the advantage. While the circular colliders (FCC-ee and CEPC) are able to complete the required runs at various luminosities faster their larger civil engineering work requires significantly more time and cost. Amongst the newer proposals only $C^3$ proposes a timescale which is suitable for early physics. The secondary consideration is the run plan, which is adjustable based on physics motivation. The ILC with the earliest start date for physics holds an advantage in spite of somewhat lower luminosity than a circular collider.
	
\item What type of running is essential and how long?
	
	With the premium placed on measuring the higgs couplings, the run at $\sqrt{s}\approx 250$ GeV is the most important. The ILC proposal is a 11-year run with various electron and positron polarization settings accumulating 2 ab$^{-1}$. The FCC-ee proposal is a 3-year run accumulating 5 ab$^{-1}$ at $\sqrt{s} = 240$ GeV using two interaction points. 
	
\item Which machines are ready to be built in 2025-2030, 2030-2035, or after 2035+ based on the technology readiness? 
	Following the ITF report Table 17, the only \ee~collider option that can deliver physics in the near-term, less than 12 years, is ILC-0.25 TeV. Midterm options, FCC-ee, CLIC and $C^3$, are estimated to take 13-18 years. Long-term options, CERC is estimated to be 19-24 years or WFAs and ERLs are estimated as $>25$ years away to first physics. The ILC could be built in 2025-2030 time scale provided the project is approved soon. $C^3$ requires a demonstrator before construction begins, so it may be compatible with beginning construction in 2030-2035. The CERN feasibility study which is underway will indicate the time early for construction to begin for FCC-ee. Seven years from the completion of the HL-LHC program is indicated as time early for FCC-ee physics operations by CERN.  
	
\item What are the costs of the various machines, also  ILC, CCC, if sited in the US. ERL is mentioned, so it would be good to see, where is it going to be sited, where in the US and where, what would be its timescale and cost? Is it as the same footing as CCC?
    
    Following the ITF report Tables 1-2 , the cost range for linear Higgs Factories (ILC-0.25, CCC-0.25 and CLIC-0.38, ReLiC-0.24) is \$7-11B while the circular Higgs factories (FCCee0.24, CEPC-0.24) have a cost range of \$12-18B.  The estimated electrical power for the linear machines is in general lower (ILC-0.25 runs at 140 MW, CCC-0.25 runs 150 MW, CLIC-0,38 runs at 170 MW) than circular machines (FCCee runs at 280 MW, CEPC runs at 340 MW). Higher energy machines of course have higher cost and electrical power requirements, ILC 500 GeV cost ranges \$11-17B and$C^3$-1000 GeV) machine is estimated between \$12-18B with power at 700 MW. The CERC (240) 600 GeV machines are estimated at (\$15-30B) \$18-32B but with reduced power needs with 90 MW at 240 GeV  .
    
    One proposed site for the $C^3$ machine is the Fermilab, although it can be built anywhere with about 8-km tunnel/cut-and-fill ability. Energy recovery linacs require low gradient operation. Therefore, ReLiC and ERLC are very large sites. The circular energy recovery based machine CERC needs 100-km scale site.
    
\item What are the primary detector R\&D needed, if a targeted Higgs Factory program is funded? 
    
    The primary detector R\&D needs concern precision timing, further reducing the material in the tracker, developing powering and cooling techniques, and making the chosen technologies scalable to a large HEP detector. 

    
\end{itemize}

\section*{Acknowledgements}

The authors thank the \ee-community for the lively discussions at the Community Summer Study at Seattle, the authors of the white papers, and the Snowmass topical forum conveners. This work is supported by the US Department of Energy. 

\printbibliography

\end{document}